\documentclass[%
 aip,
 amsmath,amssymb,
 reprint,%
]{revtex4-1}

\usepackage{graphicx}
\usepackage{dcolumn}
\usepackage{bm}
\graphicspath{{./figures/}}
\usepackage[utf8]{inputenc}
\usepackage[T1]{fontenc}
\usepackage{mathptmx}
\usepackage{subfigure}
\usepackage{xcolor}

\begin{document}

\preprint{AIP/123-QED}

\title[Role of viscoelasticity on the dynamics and aggregation of chemically active sphere-dimers]{Role of viscoelasticity on the dynamics and aggregation  of chemically active sphere-dimers\\}

\author{Soudamini Sahoo}
\email{soudamini@iiserb.ac.in}
\affiliation{Department of Physics, Indian Institute of Science Education and Research Bhopal, Bhopal, India, 462066}

\author{Sunil Pratap Singh}
\email{spsingh@iiserb.ac.in}
\affiliation{Department of Physics, Indian Institute of Science Education and Research Bhopal, Bhopal, India, 462066}

\author{Snigdha Thakur*}
\email{sthakur@iiserb.ac.in}
\affiliation{Department of Physics, Indian Institute of Science Education and Research Bhopal, Bhopal, India, 462066}

* sthakur@iiserb.ac.in



\begin{abstract}

The impact of complex media on the dynamics of active swimmers has gained a thriving interest in the research community for their prominent applications in various fields.  This paper  investigates the effect of viscoelasticity on the dynamics and aggregation of chemically powered sphere-dimers by using a coarse-grained hybrid mesoscopic simulation technique.  The sphere-dimers perform active motion by virtue of the concentration gradient around the swimmer's surface, produced by the chemical reaction at one end of the dimer. We observe that the fluid elasticity enhances translational and rotational motion of a single dimer, however for a pair of dimers, the clustering in a particular alignment is more pronounced. In case of multiple dimers, the kinetics of cluster formation along with their propulsive nature are presented in detail. The key factors  influencing the enhanced motility and the aggregation of dimers are  the concentration gradients, hydrodynamic coupling and the microstructures present in the system. 

\end{abstract}

\maketitle

%

\section{Introduction}
\label{sec:introduction}

Active swimmers have attracted tremendous interest in scientific community owing to their ability to exploit  surrounding energy for performing various biological and mechanical functions~\cite{li14,Rama_2010,marchetti2013hydrodynamics,cates2012diffusive,bechinger2016active,zottl2016emergent}.  Such systems include  inherently active units which  drive them out of equilibrium. In most  of the cases the communication between them sets up  through a number of complex signaling mechanisms like interactions among neighbors~\cite{Vicsek-1995}, hydrodynamic interactions~\cite{palacci2010sedimentation,gomez16}, chemotaxis~\cite{hong2007chemotaxis,popescu2018chemotaxis}, quorum sensing~\cite{Thomas2018self,barrios2006quorum}, etc.  A suspension of active natural entities like bacteria, spermatozoa and their synthetic analogue are known to display very intriguing phenomena like swarming~\cite{partridge1982,Lowen_2013}, pattern formation~\cite{budrene_1991,Rama_2010_nature}, vortex formation~\cite{yang2014self,riedel2005self}, surface accumulation~\cite{elgeti2009self,anand2019behavior},  phase transitions~\cite{Stark_2014_phase,marchetti2013hydrodynamics,cates2012diffusive,zottl2016emergent}, etc. Such unique properties make the individual as well as collective dynamics of active particles very different from their passive counterpart~\cite{soudamini2019,kapral2008,van2015collective,snigdha12}. Fascinated by various peculiar phenomena in these natural microswimmers, a new realm of  chemically powered synthetic motors started over a decade ago~\cite{kapral_2017_many-body,chen2016chemotactic,kapral2007chemically,chen2018chemically,kapral2008,gu2013self,yamamoto2015,paxton2004catalytic,prabha_2019_vesicle,chen_2020_chemical_wave}, where the associated spacial-and temporal-asymmetry is the origin of self-propulsion. A series of distinct schemes have been evolved to design self-propelled synthetic motors, most common among them is the phoretic mechanisms like diffusiophoresis~\cite{Anderson_1991}, thermophoresis~\cite{jiang2010active}, electrophoresis~\cite{nourhani2015self}. These systems also display nontrivial scenarios of an individual as well as collective motion similar to the natural motors~\cite{snigdha12,Kapral2018, Sen2009, Bocquet_2012,Vicsek-1995}.  

Microswimmers encounter diverse fluidic environment during their course of motion~\cite{Zhu_2020,G_Richie_2020,yuan2018recent,Elfring_2016,Prabha_2018_chem_comm,tung2017,Maldare_2016,Dipankara_2020}. The comprehensive studies on the effect of fluid reveal that the rheology of the surrounding medium can strongly affect the motility of   microsswimmers ~\cite{bechinger2016active,soudamini2019,datt2015squirming,Lauga_2007_viscoelastic}
An increase in translational velocity but a decrease in the rotational diffusion in viscoelastic fluids compared to Newtonian fluids for the {\it E.coli } has been established \cite{patteson2015running}. It was also demonstrated that the fluid elasticity of cervical mucus promotes the collective swimming of sperms, which can contribute to an enhanced probability of successful fertilisation~\cite{tung2017}.  In the context of artificial microswimmers, most of the studies so far are limited to Newtonian fluid~\cite{snigdha12,jiang2010active, prabha_2018, palacci2010sedimentation, buttinoni2012active,golestanian07,prabha_2019_vesicle,Patrick_2018}. A few recent  investigations have demonstrated the role of viscoelastic~\cite{Narinder_2019,gomez16,bechinger2016active,liu11,Saad2019}, shear thinning~\cite{gompper2011fene,datt2015squirming}, and shear thickening~\cite{qiu2014swimming} fluids on the self-propulsion. A report on diffusiophoretic active colloid in a polymeric solution demonstrates restricted active motion, {\it i.e.},  faster diffusion, lower motility, and shorter persistence length of directed motion~\cite{Saad2019}. A mesoscale hydrodynamic simulation of a spherical squirmer immersed in a solution of self-avoiding polymers shows a remarkable enhancement of the rotational diffusion of squirmer~\cite{winkler_2020}. This qualitatively agrees to the observed experiments on the enhancement of rotational diffusivity of self-propelled Janus particle in viscoelastic fluid~\cite{gomez16}. 
 
This article is an attempt to elucidate the collective behavior of an anisotropic active dimers in a complex medium. The self-propelled sphere-dimer comprises two connected spheres, where one sphere consumes fuel in the environment to generate a chemical gradient, and the other exploits this gradients to perform directed motion.  Our intention here is to investigate the role of viscoelasticity on the dynamics and aggregation of the active dimers.  We use a mesoscopic simulation technique based on explicit solvent that provides strength to probe the microstructures present in the system in great detail. This helps us to correlate the dynamics of the dimer with fluid micro-structural rearrangement in the course of its motion. The diffusiophoretically active dimers are immersed in a viscoelastic solvent, which is modeled using finite-extensible nonlinear elastic (FENE) dumbbells. 

We observe an enhancement in the translational and rotational motion of a single dimer in the viscoelastic fluid. Further, we demonstrate that the origin of such  enhancement is related to various dynamical quantities such as microstructural changes and hydrodynamic correlations present in the viscoelastic fluid. A substantial boost in the aggregation of dimers is also noticed in the viscoelastic fluid as compared to its Newtonian counterpart. Systematic studies of aggregation are carried out by varying the dimer densities.  The synergy between the dimers emerges due to a combination of interactions like chemical gradient, hydrodynamic correlations, and fluid microstructures.

The work is organized as follows. The model of a chemically active dimer along with the viscoelastic fluid and their governing equations are described in Section \ref{sec:simulation model}.  Section \ref{sec:results} presents the effect of medium viscoelasticity on the translational and rotational dynamics of a single diffusiophoretic dimer, a pair-dimer, and the clustering dynamics in case of multiple dimers. Finally, the conclusions of the study are drawn in Section \ref{conclusion}. 

\section{Simulation Model}
\label{sec:simulation model}

\begin{figure}
\includegraphics[width=0.85\linewidth]{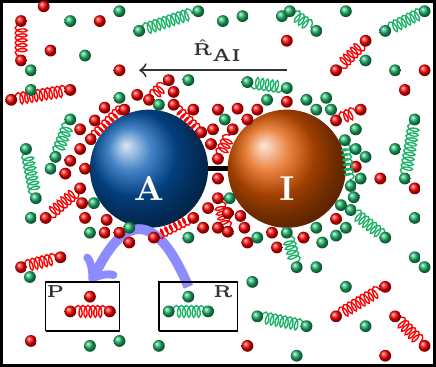}
\caption{A schematic picture includes a dimer, few dumbbells, and points like solvent particles. The two big connected spheres represent a dimer, where blue and orange spheres represent the active $(A)$ and the inert $(I)$ monomers, respectively. The green dumbbells and solvent point-like spheres represent the reactant species $(R)$ whereas the red represent the product species $(P)$.}
\label{Fig:multi_dimer_schematic}
\end{figure}

The chemically active sphere-dimer consists of  two identical spheres of  radii $\sigma$, connected  by  a harmonic  potential~\cite{soudamini2019},
$U_d=\frac{1}{2}\kappa_d(|{\bf R}_1-{\bf  R}_{2}|-l_d)^2$.  Here, $\kappa_d$, the spring constant is chosen in a way that the large fluctuations in the average bond length of the dimer ($l_d$) are suppressed. The dimer is  embedded in a coarse-grained viscoelastic fluid mixture consisting of  point-like and dumbbell-like reactants ($R$) as shown in Fig.~\ref{Fig:multi_dimer_schematic}. The dimer is chemically active by virtue of one of the spheres, labeled as `$A$' (Active), which takes part in a chemical reaction on its surface.  The sphere `$A$' catalyzes the irreversible reaction $R + A  \rightarrow P + A$, whenever a reactant `$R$' encounters `$A$' and leaves its boundary layer~\cite{kapral2008,snigdha12,prabha_2018} at $r_c = 2^{1/6}\sigma$. Note that, when the point particle of a dumbbell-like reactant interact with $A$, the entire dumbbell get converted to a product type solvent ($P$). The other sphere `$I$' (inert) is chemically neutral. A schematic picture in Fig.~\ref{Fig:multi_dimer_schematic} indicates the chemical reaction on `$A$' that generates the product `$P$' around its neighborhood. This results in an inhomogeneous distribution of `$R$' and `$P$' around the inert sphere `$I$'.

The interaction between the dimer and solvent molecules are implemented through repulsive Lennard-Jones (LJ) potential,
\begin{align}
 U_{\alpha S}(r)&=4\epsilon_{\alpha S}\Big{[}{\Big{(}\frac{\sigma}{r}\Big{)}}^{12}-{\Big{(}\frac{\sigma}{r}\Big{)}}^6+\frac{1}{4}\Big{]},\, r<r_c \\
          &=0,\,  \textrm {otherwise} \nonumber.
\end{align}

\noindent 
Here, $S = A$ or  $S=I$ and $\alpha = R$ or $\alpha = P$ to indicate different interactions between dimer and solvent.  In particular, we take $\epsilon_{RA} =  \epsilon_{PA} = \epsilon_{RI} =\epsilon \ne \epsilon_{PI} $, to characterize the energy parameters between different pairs.  This choice of the energy parameter and the asymmetry around the inert sphere is the key ingredient for the self-propulsion of the sphere-dimer.  We choose $\epsilon > \epsilon_{PI}$ to enable the dimer to move with the active sphere ($A$) at the front  as explained in section ~\ref{sec:results}. For continuous propulsion of dimer, the continuous supply of reactant `$R$' shouldn't be disrupted. Therefore, we perform a back conversion of `$P$' to `$R$',  $4\sigma$ away from every dimer's center-of-mass position, to maintain the required concentration of `$R$'.

In case of multi-dimers, the interaction between two monomers separated by a distance $r$ is implemented through repulsive Lennard-Jones (LJ) potential
with interaction energy $\epsilon_d$ and diameter $2\sigma$. The motion of a sphere-dimer and fluid particles are updated by hybrid molecular-dynamics multi-particles collision (MD-MPC) technique~\cite{kap2008}.
  
The fluid is modelled in the frame work of MPC dynamics~\cite{kap2008,gompper2009mpcd}, where usually the fluid particles are considered as point particles. To bring viscoelasticity in the medium, we connect a few pairs of point-like solvent molecules through a finite-extensible nonlinear elastic (FENE) potential~\cite{gompper2011fene} to make them dumbbell-like,
\begin{equation}
U(\textbf{r}_i,\textbf{r}_{i+1}) = -\frac{\kappa}{2}{R_0}^2\ln \left[ 1-{ \left(\frac {\textbf{r}_i-\textbf{r}_{i+1}}    {R_0}  \right) }^2  \right].
\end{equation}
Here, $\kappa$ is the spring constant and $R_0$ is the maximum  extension of the spring.  We consider total $N_s$ solvent point particles in solution, out of which $N_b$ forms  the dumbbells.  The extent of viscoelasticity in the medium is altered by varying $N_b=0$ to $N_s$.  $N_b=0$ corresponds to only point-like particles that makes  Newtonian fluid (marked as $0\%$ in figures), whereas $N_b = N_s$ have only  dumbbell-like particles resulting in a  viscoelastic fluid (read as $100\%$ in figures). We also consider the fluid to be a mixture of point-like and dumbbell-like particles. We define ${f_d} = ({N_b}/{N_s})\times 100$ to represent the percentage of point-like particles which form the dumbbells. 


The positions and velocities of the solvent particles are updated by the MPC rules, which involve streaming and collision steps. The streaming step performed at every MD time ($\Delta t$) comprises updating the positions and velocities of solvent via Newton's equation of motion. The collision step is carried out at a time interval of $\tau$. Here all the solvent particles are sorted into a grid of cubic cells of size $a$ where they interact with other molecules. The relative velocities of  the particles  $(\delta {\bf{v}}_i= {\bf{v}}_i-{\bf{v}}_{cm})$  with respect to the centre-of-mass velocity $({\bf{v}}_{cm})$ of the cell are rotated about a random axis by an angle $\gamma$. The velocity of the $i^{th}$ particle after collision is given as ${\bf{v}}_{i}(t+\tau)= {\bf{v}}_{cm}(t)+ \Re(\gamma)\delta {\bf{v}}_i$, where $\Re(\gamma)$ is the rotation matrix. Random grid shifting is performed before the collision step to ensure the Galilean invariance when mean free path of the solvent particles is smaller than the MPC cell size~\cite{ihle2001stochastic}. 
 
{\it Simulation parameters:} The physical quantities are reported in units of thermal energy $k_BT$, mass of the solvent $m$, and  the collision cell length $a$.  Time is scaled in units of $\tau_m=\sqrt{{ma^2}/{K_BT}}$,  velocity in units of $a/\tau_m$, the spring constants $\kappa$, and $\kappa_d$ are in units of ${k_BT}/{a^2}$. The  MD integration is performed at a fixed time interval, $\Delta t = 2 \times 10^{-3}$. Other simulation parameters are, radii of the dimer spheres $\sigma = 2.0$,  their masses $m_d \approx 320$ (ensures the density matching with solvent), bond length of the dimer $l_d = 4.78$, and spring constant $k_d = 5000$. We set $\epsilon_{RA} = \epsilon_{PA} = \epsilon_{RI} = k_BT$, $\epsilon_{PI} = 0.01k_BT$ and $\epsilon_d = 5k_BT$.  The MPC simulation parameters are,  rotation angle, $\gamma=130^\circ$, the average number density of the fluid $\rho=10/a^3$, and  the collision time $\tau$ is varied in the range of $10^{-1}$ to $10^{-3}$ to ensure viscosity matching of different types of fluids. Unless mentioned, the  zero-shear viscosity of the fluid is $\eta_s\approx30$. All the simulations are performed in the cubical box of dimension  $L_s = 40.0$ with periodic boundary conditions in all directions. We chose FENE parameters as $R_0=20$  and $\kappa = 0.5$ resulting in the average bond length of the dumbbells to be $2.25a$. Results have been averaged over more than 30 independent ensembles. The Reynolds number $(Re)$  for the fluid lies in the range of $10^{-3}$  to  $10^{-1}$,  similarly, Schmidt number $(Sc)$ is in the range of $10$ to $10^2$ ensuring the fluidic behavior ~\cite{padding2006}. We have considered three different concentration of the dimers with $N_d= 1, 2$ and $28$ which results  a  packing fraction range of $(10^{-3},10^{-2})$. In few results, the time is scaled as $t/\tau_s$ where  $\tau_s$ $(= L_d / {\langle V_p \rangle})$ is the time taken by the dimer to travel it's own length ($L_d \approx 9.0$) in case of a Newtonian fluid. 

\section{Results}
\label{sec:results}

The solvent properties are gradually varied from Newtonian to viscoelastic by changing the fraction of dumbbell-like solvent particle while maintaining a nearly constant viscosity. The physical quantities like translational, rotational motion of the dimers and solvent density profile  contribute to a great extent to our understanding of their dynamics and assembly. 

\subsection*{Transport properties of fluid}
\label{subsec:fluid_properties}

Before exploring the dynamics of dimers in viscoelastic fluids, here we discuss some of the essential properties of the fluid consisting of FENE dumbbells. We consider fluids having different percentage of dumbbells and to maintain their zero-shear viscosity, we tune the MPC collision time ($\tau$). The zero-shear viscosity $\eta_s$ is computed from the non-equilibrium simulations by imposing linear shear flow where the  stress tensor varies linearly with  viscosity in linear-response regime ($\eta_s = \sigma_{xy}/\dot{\gamma}$)~\cite{winkler2009stress,gompper2011fene}.  This is calculated by using Lees-Edwards boundary condition~\cite{Allen_1989} for Newtonian and different viscoelastic systems where the applied shear rate is $\dot{\gamma} = 0.005$. Table~\ref{table:viscosity} shows the choice of the parameter $\tau$ for Newtonian and various viscoelastic fluids which results in $\eta_s=30$.

 
\begin{table}[htbp]
  \caption {A constant shear viscosity $\eta_s=30$ is maintained, while varying concentrations of dumbbells  ($f_d$), at the following choice of fluid parameters.}
  \label{table:viscosity}
\begin{tabular}{ | m{1.2cm} |  m{1.2cm} | m{1.2cm} | m{1.2cm}| m{1.2cm}| m{1.2cm}|}
\hline
$~~f_d$  & ~~0   & ~~10 &  ~~30 &  ~~50 & ~~100 \\
\hline
$~~\tau$ & ~~0.028 & ~~0.042  & ~~0.072 &  ~~0.104 & ~~0.18 \\ \hline
\end{tabular}
\end{table}

To quantify the elastic strength of the fluids for  different $f_d$, we have computed the variation of storage modulus ($G^\prime$) as a function of $f_d$. This is performed by imposing an oscillatory strain as, ${\gamma} = \gamma_0 sin(\omega t)$ where $\gamma_0$ is the strain amplitude and $\omega$ is frequency of oscillation. The storage modulus $(G^\prime)$ and loss modulus $(G^{\prime\prime})$ can be measured in the simulations by following the stress-tensor expression~\cite{tao2008}

\begin{equation}\label{eq:stor-loss}
\sigma_{xy}(t) = \gamma_0 [G^\prime(\omega) \sin(\omega t) + G^{\prime\prime} (\omega) \cos(\omega t)].
\end{equation}

\begin{figure}[h]
\includegraphics[width=\linewidth]{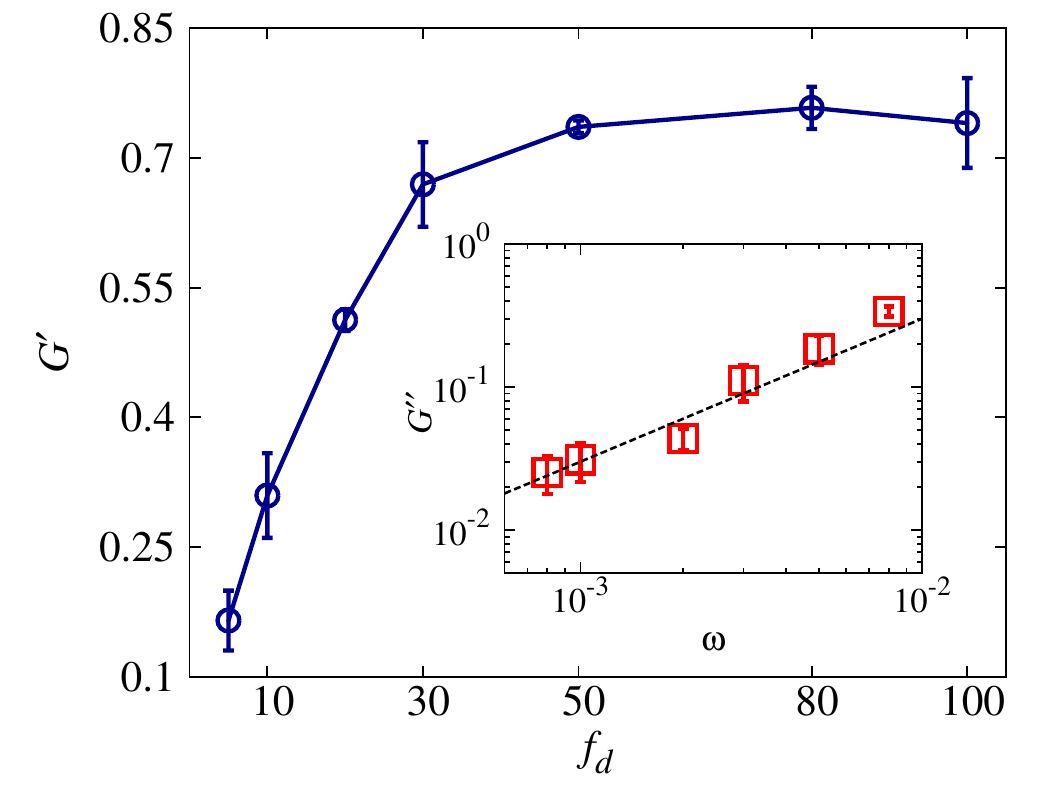}
\caption{Variation of storage modulus $(G^{\prime})$ for viscoelastic fluids having different $f_d$ with $\gamma_0$ = 0.2 and $\omega$ = 0.1. Inset: Loss modulus $(G^{\prime\prime})$ as a function of $\omega$ for a viscoelastic fluid ($f_d = 100$) with $\gamma_0$ = 0.2. The black dashed line is a fit of $G^{\prime\prime} = \eta_s \omega$, $\eta_s=30$.}
\label{Fig:stor_loss}
\end{figure}

The storage modulus is identified at $\sin(\omega t) = \pm 1$  and similarly the loss modulus is obtained  at $cos(\omega t) = \pm 1$. The measured values are shown in Fig.~\ref{Fig:stor_loss}, where the storage modulus initially increases with $f_d$ following a saturation for $f_d \ge 50$.  The contribution of this elasticity reflects in the  dynamics and aggregations of the dimers, which is our prime focus. The inset shows the variation of storage modulus ($G^{\prime\prime})$ with $\omega$. As expected in regime of  $\omega < 0.01$, $G^{\prime\prime}$ follow a linear behavior and the viscosity calculated in this region ($\eta_s = \frac{G^{\prime\prime} (\omega)}{\omega} \approx 30$) which is same as that obtained from the linear shear-flow. For the calculation of $G^{\prime}$ and $G^{\prime \prime}$, 40 or more cycles are considered and for each ensemble for better statistics.

\subsection*{Case I: Single active dimer}
\label{subsec:1_dimer}

The self-propulsion of chemically active sphere-dimer has been extensively studied in the context of Newtonian fluid~\cite{prabha_2018,kapral2008,snigdha12,colberg_kapral_2014}. The chemically active sphere `$A$' converts reactants `$R$' species to the product `$P$', thereby generating a  symmetric distribution of `$R$' and `$P$' species around `$A$'. However, the diffusion of solvent species produces a concentration asymmetry of `$R$' and `$P$' in the vicinity of inert sphere `$I$' (see Fig.~\ref{Fig:multi_dimer_schematic}). The combination of non-equilibrium concentration gradient and the interaction potential difference of `$R$' and `$P$' with `$I$' causes the directed motion of the dimer, which is also known as self-diffusiophoresis~\cite{Anderson_1991,kapral2008}.  The direction of the dimer's motion depend on the choice of  $\epsilon {\rm{~and~}}\epsilon_{PI}$. For $\epsilon > \epsilon_{PI}$, the inert sphere experience a net force from the solvent in a direction of $\hat{R}_{AI}$, therefore it moves with $A$ sphere at the front end. In this part, our goal is to highlight the effect of viscoelasticity on the propulsion dynamics of a single  dimer.

\begin{figure}[h]
\includegraphics[width=\linewidth]{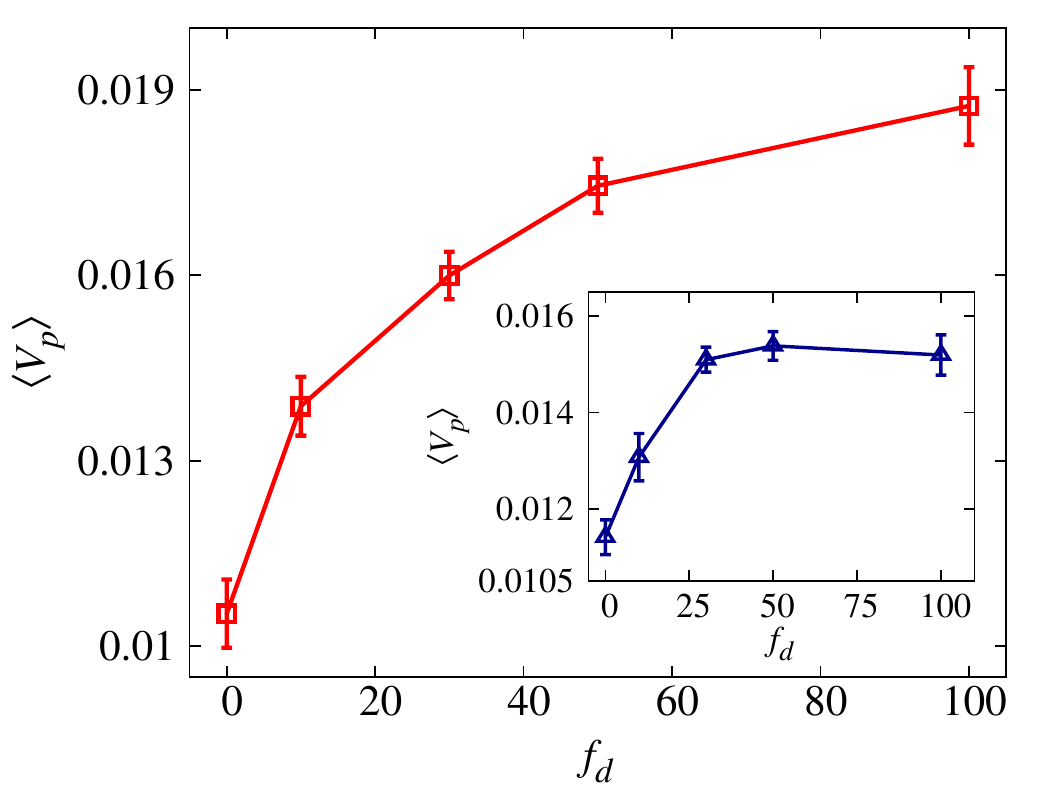}
\caption{The average directed velocity $\langle V_P \rangle$ of the dimer is plotted for various $f_d$ when the entire reactant dumbbell take part in the chemical interaction with `$A$' . The inset shows the same but when only the reactant molecule linked to the dumbbell interact with `$A$' and converts to `$P$'.}
\label{Fig:1_dimer_vz}
\end{figure}

To quantify the translational motion of a single dimer, we calculate directed speed of the dimer  along its axis $(\hat{R}_{AI})$,  defined as $V_p = {\bf V}_{cmv}\cdot  \hat{R}_{AI}$ where  ${\bf V}_{cmv}$ is center-of-mass velocity.
Figure~\ref{Fig:1_dimer_vz} shows the variation of $\langle V_P \rangle$ with increasing dumbbell percentage ($f_d$). The averaging ($ \langle ~~ \rangle$) is  performed over  time and ensembles. It is quite evident that the viscoelasticity of the medium  enhances the average directed velocity of the dimer to approximately twice as compared to the Newtonian fluid. Moreover the directed velocity exhibits a non-linear increase with the elasticity of the medium.


 As mentioned in the section~\ref{sec:simulation model}, when a reactant point particle linked with a dumbbell interact with `$A$', the entire dumbbell get converted to `$P$' type solvent. Hence, it is expected to have higher reaction rate  with increase of $f_d$. To observe the solo effect of the viscoelasticity over simple fluid, we compute  directed speed  for the same reaction rate. Here, the chemical reaction converts  to `$P$' only that part of dumbbell  which physically interacts with `$A$'.  We found directed speed $\langle V_P \rangle = 0.0155$ for a viscoelastic fluid at $f_d = 100$ which is  nearly $50\%$ increase in $\langle V_P \rangle$ compared to the Newtonian fluid for the same viscosity (see inset of Fig~\ref{Fig:1_dimer_vz}). This clearly indicates the effect of elasticity on  the enhancement of  directed speed of the dimer. Henceforth, for the simplicity  we  will stick to the case where the entire dumbbell is involved in the chemical reaction (see Fig.~\ref{Fig:multi_dimer_schematic}).

\begin{figure}[t]
\includegraphics[trim={5.2cm 9.7cm  5.2cm  9.85cm},clip,width=\linewidth]{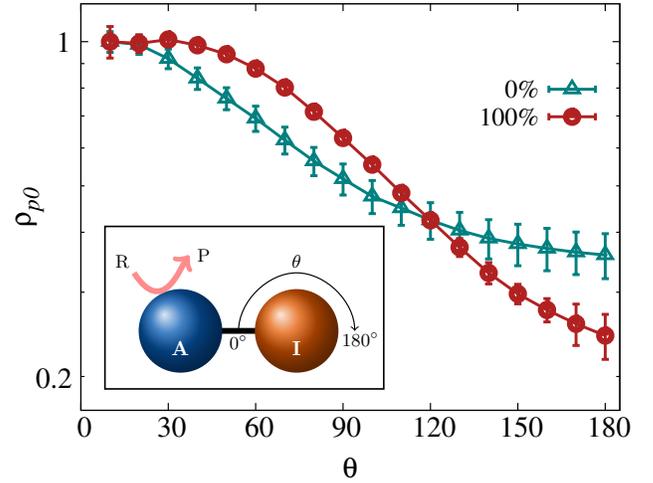}
\caption{The normalised density of  $P$ around the inert sphere for different kinds of fluid (semi-log in y-axis). The inset shows a schematic picture defining $\theta$.}
\label{Fig:1_dimer_conc_grad}
\end{figure}

It is established  that for diffusophoertic motion, the concentration gradient of the fluid around the dimer plays a key role in dictating its propulsion behavior ~\cite{Anderson_1991}.
 Therefore, to comprehend the translational motion of the dimer, we compute the  concentration gradient of product  species ($P$) around inert sphere ($I$). Figure~\ref{Fig:1_dimer_conc_grad} shows the normalised local density, $\rho_{p0}$ = $\rho_p$/$\rho_0$ ($\rho_0$ being the bulk density) of $P$ around the inert sphere. For $\theta \geq 90^{\circ}$, the variation in $\rho_{p0}$ with $\theta$ exhibits a faster decay for  $f_d=100$, emphasizing a stronger  concentration asymmetry in the case of viscoelastic fluid. 

\begin{figure}[h]
\includegraphics[width=0.93\linewidth]{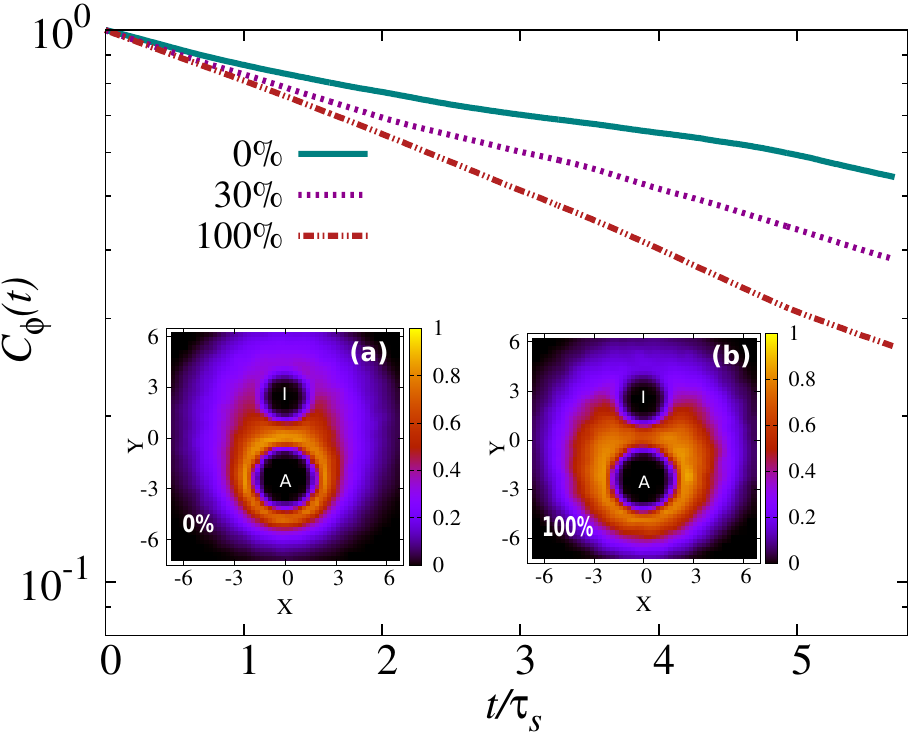}
\caption{The relaxation of orientation vector as its  auto-correlation function ($C_\phi$(t)) of the dimer in different kinds of fluid with viscosity $\eta_s=60$ (semi-log in y-axis). Inset: The density  of $P$ solvents around a dimer in a plane of thickness $2a$ passing through the dimer. }
\label{Fig:1_dimer_rotation_colormap}
\end{figure}

To understand this further, we show a color-map of the normalized local product density $\rho_{p0}$ in inset of Fig.~\ref{Fig:1_dimer_rotation_colormap}. The variation of $\rho_{p0}$ in a XY plane of thickness $2a$ passing through the center of the dimer is shown in the color-map. A higher degree of asymmetry around the dimer, in case of viscoelastic fluid, is obvious from the colormap.  Important to note here that viscoelastic medium has a steeper gradient around inert sphere ($I$), which might be responsible for the enhanced directed velocity. 
Similar microstructural rearrangements of fluid elements has also  been addressed in  our previous work~\cite{soudamini2019}. 

The local environment may impact the rotational motion of the dimer along with translational motion. To quantify the rotational motion, we calculate the orientational correlation function, $C_\phi(t) = \langle \hat R_{AI}(t) \cdot \hat R_{AI}(0) \rangle $. Figure~\ref{Fig:1_dimer_rotation_colormap} compares  $C_\phi(t)$ of the dimer for various $f_d$. 
It is evident from the correlation that the rotational dynamics of dimer is faster with the increasing degree of elasticity in the medium. The orientational relaxation time $\tau_R$,  obtained from the exponential decay of $C_\phi(t) \sim  e^{-\frac{t}{\tau_R}}$, is shown in Table~\ref{tau_per}. Such enhancement in rotational dynamics of active particles in viscoelastic fluid has also been observed in recent studies~\cite{gomez16,winkler_2020}.

\begin{table}[h]
  \caption{\ The scaled orientational relaxation time $\tau_R/\tau_s$  with $f_d$.}
  \label{tau_per}
\begin{tabular}{ | m{1.5cm} |  m{1.5cm} | m{1.5cm} | m{1.5cm}|}
\hline
$~~~f_d$  & ~~~0   & ~~~30 & ~~~100 \\
\hline
$~~~\tau_R/\tau_s$ & ~~~7.5 &  ~~~5.5 & ~~~4.5\\ 
\hline
\end{tabular}
\end{table}

\subsection*{Case II: A pair of active dimers}
\label{subsec:2_dimer}

\begin{figure}[t]
\includegraphics[trim={0.0cm 0.0cm  0.0cm  -0.27cm},clip,width=0.95\linewidth]{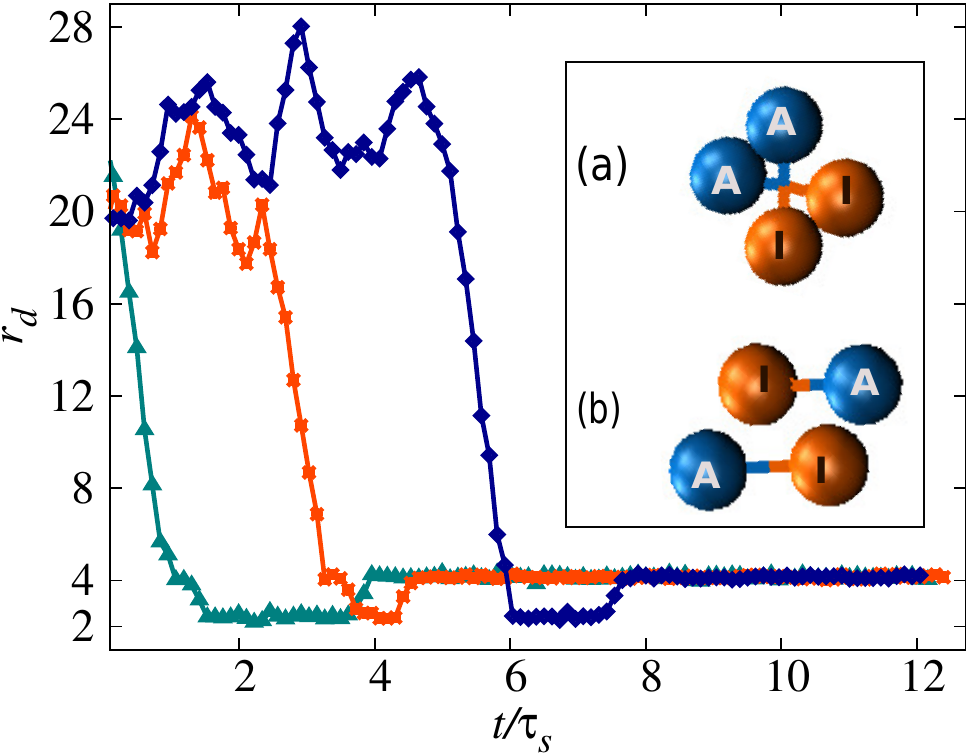}
\caption{Variation of the distance between two dimer's center-of-mass position ($r_d$) with time for three different simulation realisations. The inset shows the observed self-assembled configurations in case of a pair of dimers in  the viscoelastic fluid.}
\label{Fig:2_dimer_snap_shots}
\end{figure}

We intend to present a detailed study of the collective dynamics of chemically active dimers in viscoelastic medium. Before attempting the multi-dimer case, we first report the characteristic dynamics of a pair of dimers, which will enlighten our understanding of collective dynamics.  The inter-dimer center-of-mass position separation ($r_d$) with time  is measured in Fig.~\ref{Fig:2_dimer_snap_shots} for a viscoelastic fluid with $f_d=100$. As seen, at long time, dimers are likely to stable at $r_d\approx4$ and this separation distance mostly corresponds to  anti-parallel configurations as shown in inset (b).  Note that before achieving the stable anti-parallel configuration a short-lived cross configuration (inset (a)) with $r_d<3$ is observed in the system.

\begin{figure}[h]
\includegraphics[trim={0.0cm 0.0cm  0.0cm  0.3cm},clip,width=1.05\linewidth]{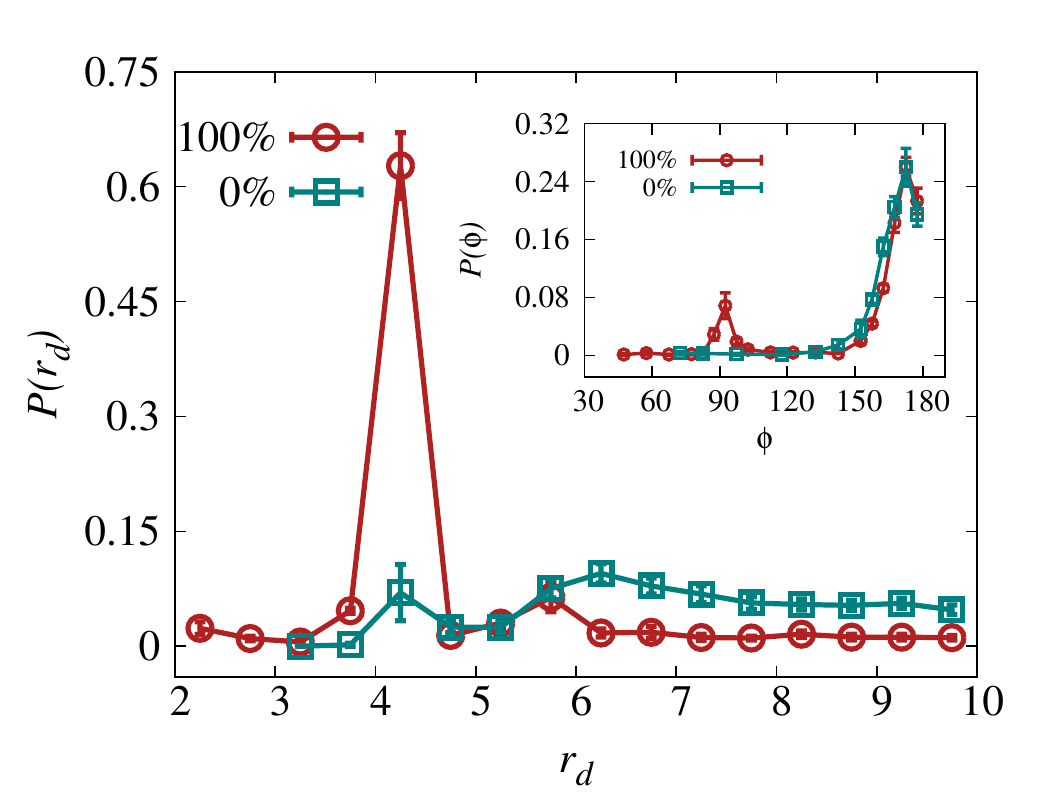}
\caption{The probability distribution function of dimer's separation $r_d$ for both Newtonian and viscoelastic fluids. The inset contains the plot for the probability distribution of angle between the dimer's axis ($\phi$) for $r_d \le 5a$.}
\label{Fig:2_dimer_dist_ang}
\end{figure}

To substantiate the configurations observed in Fig.~\ref{Fig:2_dimer_snap_shots}, we quantify the spatial arrangements of the active dimers with the help of average inter-dimer center-of-mass separation ($r_d$) and alignment angle ($\phi$). The probability distribution $P(r_d)$ for the dimers to be  at a distance $r_d$ for both Newtonian ($0 \%$) and viscoelastic ($100 \%$) fluid is shown in Fig.~\ref{Fig:2_dimer_dist_ang}. A peak in $P(r_d)$ at $r_d \approx 4 = 2\sigma$ is observed in both cases. This peak corresponds to the structure, where the center of the two dimers are at their diameter distance apart like in the inset (b) of Fig.~\ref{Fig:2_dimer_snap_shots}. Notice that the height of the peak is much stronger in case of the viscoelastic medium. Further, the Newtonian fluid doesn't have any cross configurations with $r_d < 3.0$, whereas its viscoelastic counterpart does possess such configurations. As seen from Fig.~\ref{Fig:2_dimer_snap_shots},  the observed cross configurations  in viscoelastic fluid is a short-lived which finally leads to a stable anti-parallel configuration at $r_d\approx 4$.

The local ordering of dimers, quantified with an average  angle $\phi$ between them when $r_d \le 5$, provides more insight into these configurations. The inset of Fig.~\ref{Fig:2_dimer_dist_ang} illustrates the probability distribution ($P(\phi)$) of  angle $\phi$.
Two peaks in the viscoelastic fluid at $\phi \approx 175^\circ$ and $ 90^\circ $, indicate the preference of dimer pair to be in an anti-parallel and cross configurations, respectively, substantiating the main plot of Fig.~\ref{Fig:2_dimer_dist_ang}.
The strong peaks at $\phi \approx 175^\circ$ indicate that the dimer pair favors the anti-parallel configuration for both  types of fluid. Such anti-parallel alignment can be interpreted in terms of the effective attractive interaction between the active sphere ($A$) of one dimer with the inert sphere ($I$) of other, mediated by the product particles~\cite{Prabha_2018_chem_comm}.
This effective attraction arises due to the choice of $\epsilon > \epsilon_{PI}$. In a system of multiple dimers an inert sphere will always get influenced by the presence of a chemical gradient irrespective of which active sphere produces it. Therefore, the $I$ spheres gets influenced by both $A$s simultaneously. Which results in a net attraction to the two dimer and hence an anti-parallel configuration.

The distribution of $\phi$ confirms the absence of  intermediate cross configuration in the Newtonian fluid.  The above results demonstrates that the presence of elasticity in the medium modifies the effective interactions leading to a stronger self-assembly. 

\begin{figure}[h]
\includegraphics[width=0.95\linewidth]{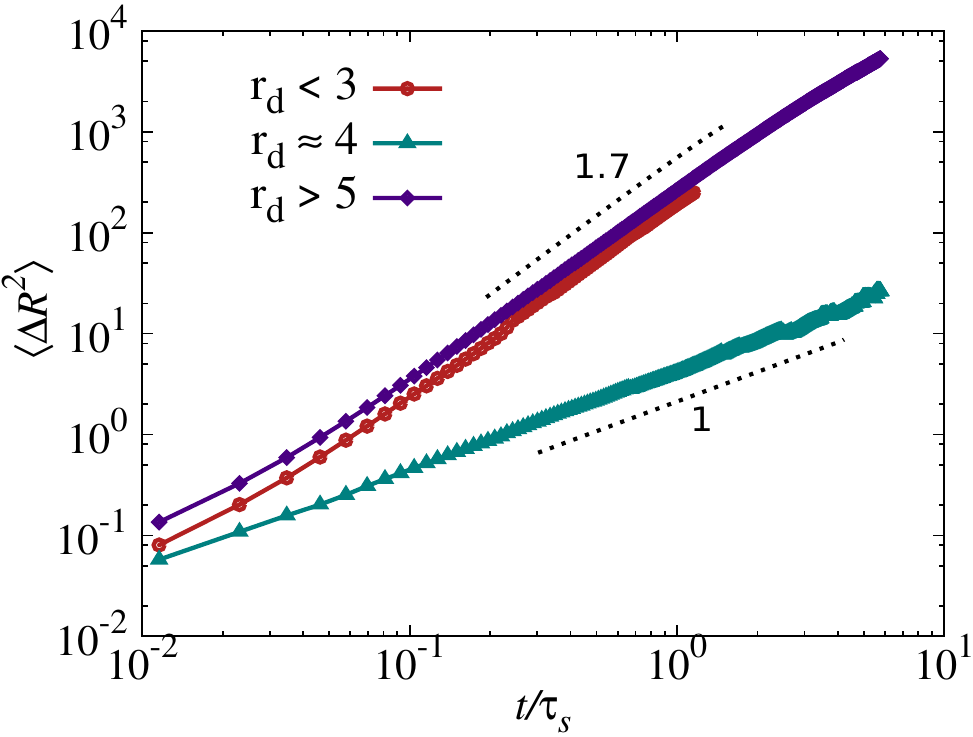}
\caption{The average mean-squared displacement of a dimer in  different structures of cluster in viscoelastic fluid.}
\label{Fig:2_dimer_msd}
\end{figure}

Next, we focus on the dynamical behavior of the active pair-dimer. An interesting dynamical feature due to the correlation between the dimer pair is captured in the mean-squared-displacement (MSD),
$\langle {\Delta {R}}^ 2 (t) \rangle = \langle ({R}(t) - {R}(0))^2 \rangle $, where ${R}(t)$ is the centre-of-mass position of the dimer at time $t$. Typically, $\langle {\Delta {R}}^2(t) \rangle  \propto t^\beta$ where
$\beta = 1$ for diffusive motion and  $\beta > 1$ represents superdiffusive motion. For a free active dimer case, it has been shown  that  $\beta>1$ in the intermediate time-regime~\cite{snigdha12,Golestanian_2009}.  Figure~\ref{Fig:2_dimer_msd} presents MSD  for a cross configuration ($r_d<3$), an anti-parallel assembly ($r_d\approx 4$), and for a free dimer ($r_d>5$). We observe $\beta > 1$ for both $r_d>5$  (free dimer) and $r_d<3$ (cross-configurations) indicating their propulsive nature. Surprisingly, in the anti-parallel assembly ($r_d \approx 4$) the transport of dimer pair exhibit a diffusive behavior with $\beta = 1$. 

\begin{figure}[htbp]
\includegraphics[width=0.95\linewidth]{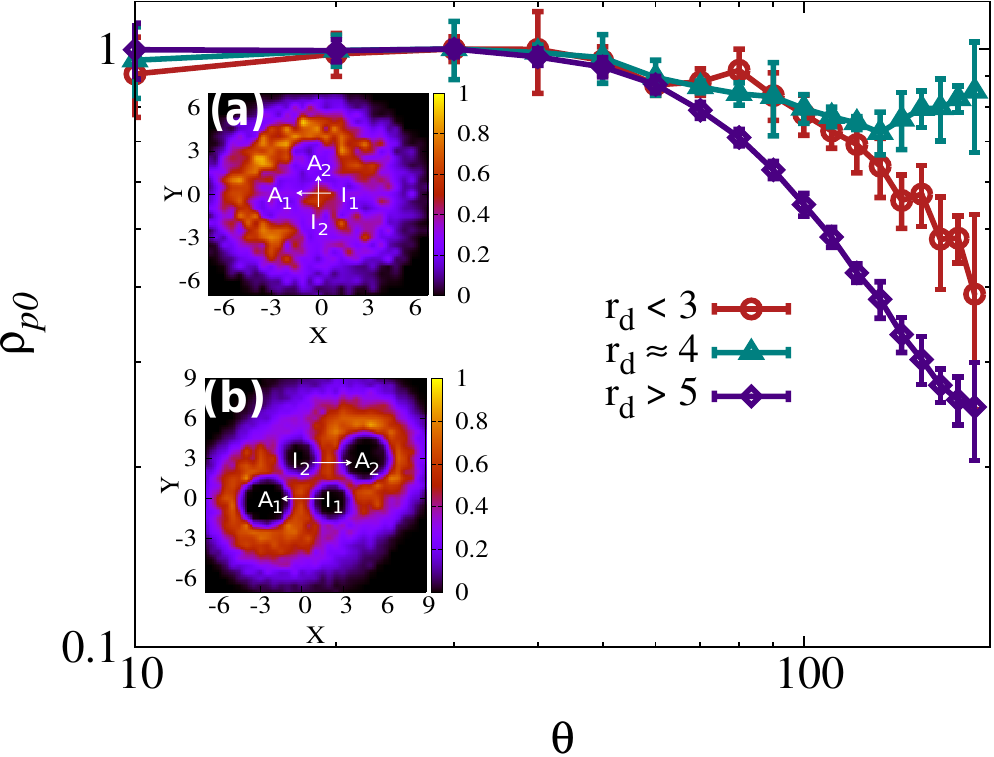}
\caption{The normalised density $\rho_{p0}$ of $P$ around the inert sphere for different spatial distances in the viscoelastic fluid. Inset: The color map for the density  distribution of product particles around the dimers for viscoelastic fluid. (a) Top view of a cross aggregation, (b) anti-parallel structure.
}
\label{Fig:2_dimer_conc_grad_color_map}
\end{figure}

To unravel the MSD behavior of pair assembly, we take a closer look at the density distribution of the product $P$ around the pair. It is quite evident from the color-map (inset of Fig.~\ref{Fig:2_dimer_conc_grad_color_map}) that the asymmetry of the product ($P$), essential for the propulsion of dimers is lost as soon as the dimer-pair assemble in an anti-parallel configuration, leading to their diffusive behavior. However, the cross configurations are able to maintain the product ($P$) asymmetry (due to the arrangement of chemically active sphere ($A$) on one side) and, hence exhibit directed motion. This color-map  further validates by the mean normalized product density ($\rho_{p0}$) in the vicinity of inert sphere ($I$) (see Fig.~\ref{Fig:2_dimer_conc_grad_color_map}). It is  apparent that for anti-parallel configuration ($r_d \approx 4$), concentration gradient of $P$ diminishes due to the presence of other active sphere ($A$).

To summarize, the concentration gradient of product particles facilitates a long-range attraction among the inert and active spheres of different dimers. When the dimers come together by long-range attraction, they  arrange themselves in their favorable  anti-parallel configurations.  However, in some cases before achieving the anti-parallel configuration a short-lived cross configuration is also observed in the viscoelastic fluid.

 \subsection*{Case III: Collection of dimers}
\label{subsec:28_dimer}

\begin{figure}[t]
      \includegraphics[width=\linewidth]{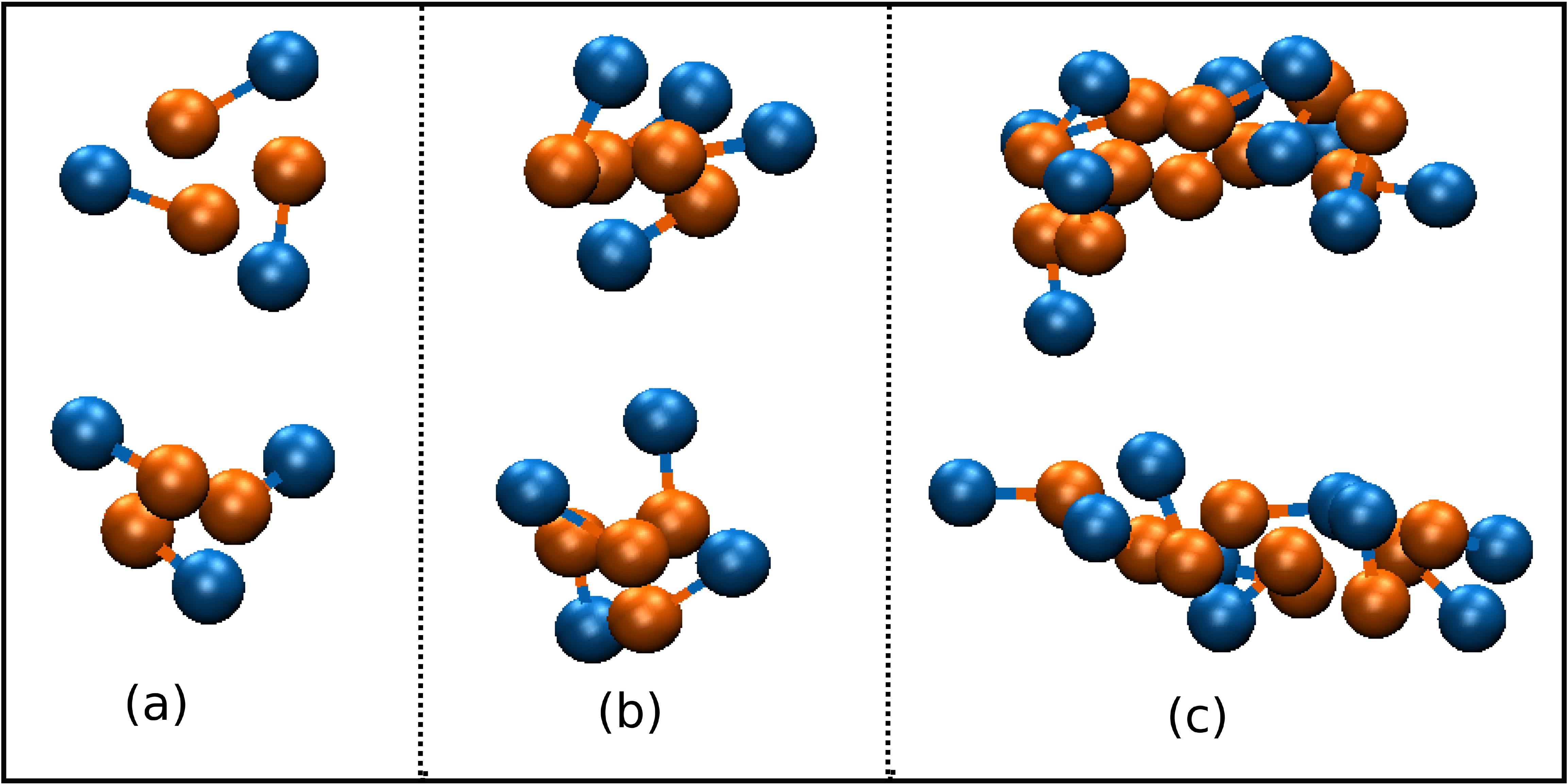}
     \caption{A few representative clusters of (a) 3 dimers, (b) 4 dimers and (c)  more than 6 dimers. Blue sphere- Active sphere, Orange sphere- Inert sphere.}
     \label{Fig:28_dimer_snap}
\end{figure}

 The obvious extension of the work is the investigation of collective dynamics in both media. For this, we  consider $N_d=28$ corresponding to a dilute solution of  packing fraction $0.065$. As discussed in the case of a pair-dimer, the long-range attraction between the active ($A$) and inert ($I$) spheres of different dimers results  formation of clusters. Similar interaction drives the clustering  in the multiple dimers (see Fig.~\ref{Fig:28_dimer_snap} a-c). 
 There are two noteworthy points here. Firstly, the inert spheres ($I$) are attracted towards active spheres ($A$) of other dimers by virtue of  the concentration field of `$P$'. And secondly, the individual dimer likes to maintain it's propulsion by keeping its `$A$' sphere at the front while moving. The competition between these two factors results in aggregates  where the  active  spheres ($A$) are often outward. Such structures are very dynamic and are quantified in terms of the distribution of cluster size and number of clusters. 

\begin{figure}[h]
      \includegraphics[width=\linewidth]{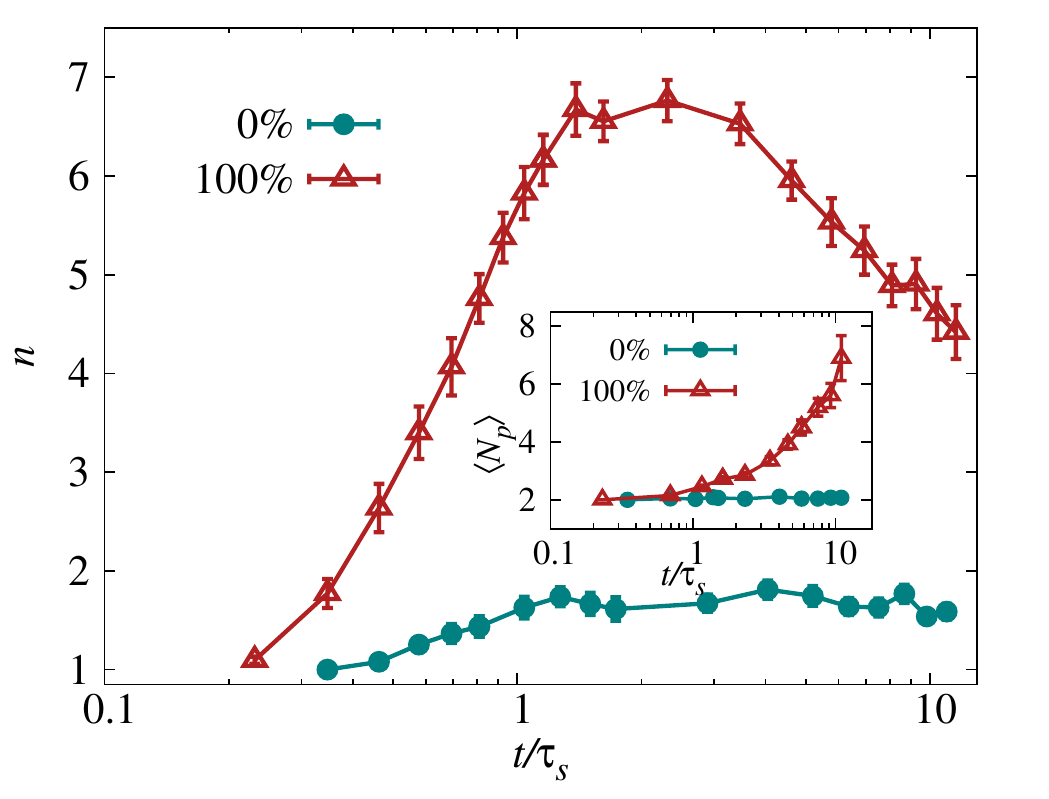}
     \caption{The kinetics of total number of clusters ($n$) as function of scaled time $t/ \tau_s$. The inset shows variation of average cluster size ($\langle N_p \rangle$) with time at $f_d=0\%$ and $100\%$. }
      \label{Fig:28_dimer_cluster}
\end{figure}

To quantify the aggregation, its kinetics is monitored by computing  total number of clusters `$n$'  over time ($t$). A dimer is part of a cluster if it falls within a cut-off distance $r_c=4.5$ from another dimer already present in the same cluster.  A non-monotonic behavior of $n$  as a function of time for the viscoelastic fluid is revealed in  Fig.~\ref{Fig:28_dimer_cluster}. Initial increase of the number of clusters  ($n$)  is by virtue of the long-range attraction among the dimers which is further followed by  the coalescence of clusters resulting in the decrease of `$n$' in the long-time limit. In addition, we calculate the average number of dimers in a cluster $\langle N_p \rangle$, defined as $\langle N_p \rangle = \langle N_m/n \rangle $ where $N_m$ is the total number of dimers that are part of any cluster.

For the viscoelastic fluid, the inset of Fig.~\ref{Fig:28_dimer_cluster} shows an increase in $\langle N_p \rangle$ for $\frac{t}{\tau_s}>1$. However, before that, the $\langle N_p \rangle$ remains roughly at two, indicating that even though the number of clusters in the system `$n$' is increasing, most of them are clusters with dimer pair. Beyond this time, the increase in  $\langle N_p \rangle$, along with the dynamics of `$n$' implies the evolution of larger clusters. In contrast to this, the plateau in both `$n$' and $\langle N_p \rangle$ for the Newtonian fluid implies the existence of dimer pairs only for the mentioned simulation range.

\begin{figure}[b]
     \centering
   \includegraphics[width=\linewidth]{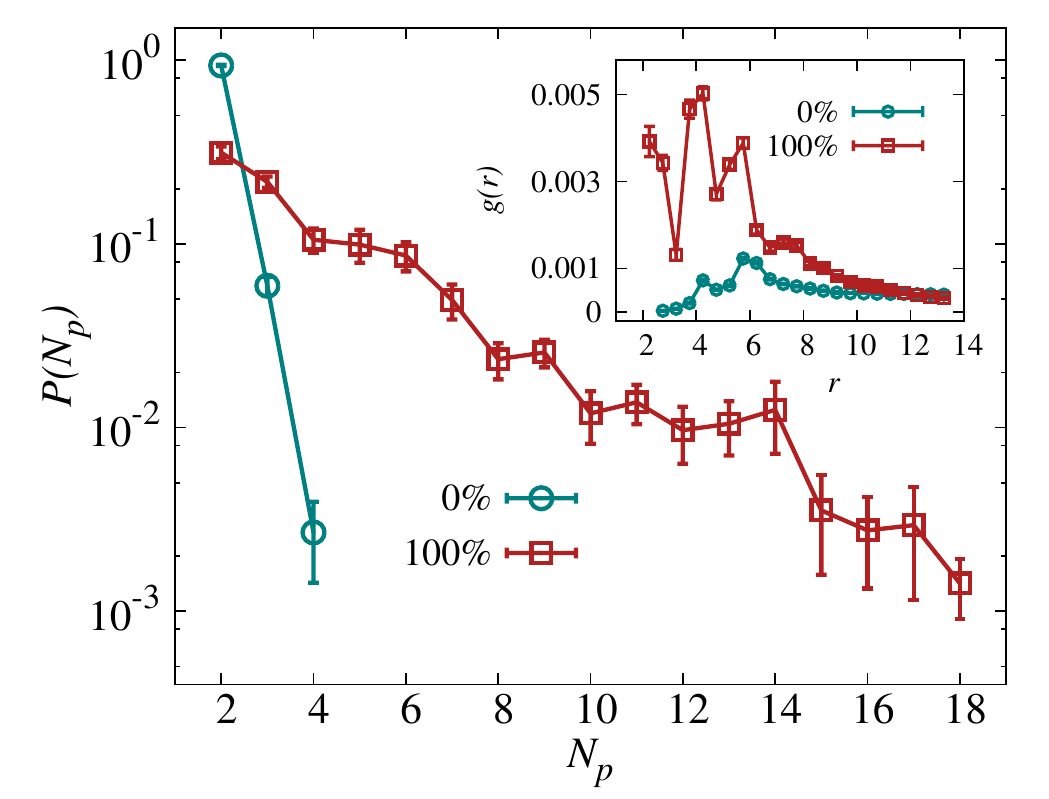}
     \caption{The probability distribution of different cluster sizes in  viscoealstic  and Newtonian fluid.  The inset shows the radial distribution function ($g(r)$) of system for both kinds of fluid.}
     \label{Fig:28_dimer_gr}
\end{figure}
To probe the pronounced clustering of dimers, we calculate the probability of finding a cluster of size $N_p$.  It is clear from  Fig.~\ref{Fig:28_dimer_gr} that for the viscoelastic fluid larger clusters exists in the system with substantially higher probability, which is not the case for Newtonian fluid even-though  the fluids are at viscosity.


Further, we quantify the spatial correlation of dimers by calculating the radial distribution function~\cite{Rapaport2004} defined as,
\begin{equation}
g(r) = \frac{1}{{\cal V}(r)} \Big \langle   \sum_{i \neq j=1}^{N_d} \delta (r - r_{ij}) \  \Big \rangle .
\end{equation}
Here $r_{ij}$ is the  distance between  center-of-mass of dimers and ${\cal V}(r)$ is the volume of the thin concentric spherical shell of inner radius $r$ and outer radius $r+\delta r$ where $\delta r$ is $0.5a$. The radial distribution function displays various peaks, specifically for the viscoelastic fluid attributing to  the presence of large clusters as illustrated in inset of  Fig.\ref{Fig:28_dimer_gr}. Therefore, the observations related to `$n$', $P(N_p)$, and $g(r)$ indicate that the elastic nature of the fluid promotes aggregation of active dimers.

\begin{figure}[h]
 \centering
     \includegraphics[width=0.95\linewidth]{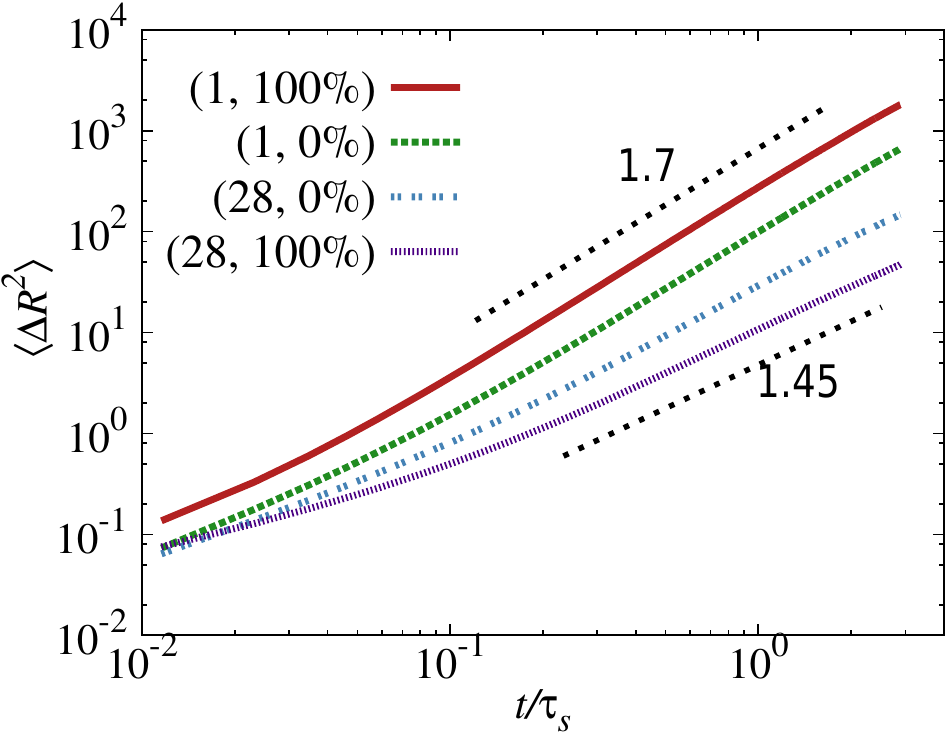}     
   \caption{The MSD of a dimer with $N_d = 1$ and  $N_d=28$  in both kinds of fluid. The index ($N_d$,$f_{d}$) represent the number of dimers $(N_d)$ and the percentage of the solvent used for viscoelastic fluid elements ($f_{d}$) in the system. The solid-lines illustrates the superdiffusive behavior with  exponent $\beta=1.45$ and $1.7$ for $N_d=28$ and $N_d=1$, respectively.}
     \label{Fig:28_dimer_msd}
\end{figure}

\begin{figure}[h]
 \centering
      \includegraphics[width=0.93\linewidth]{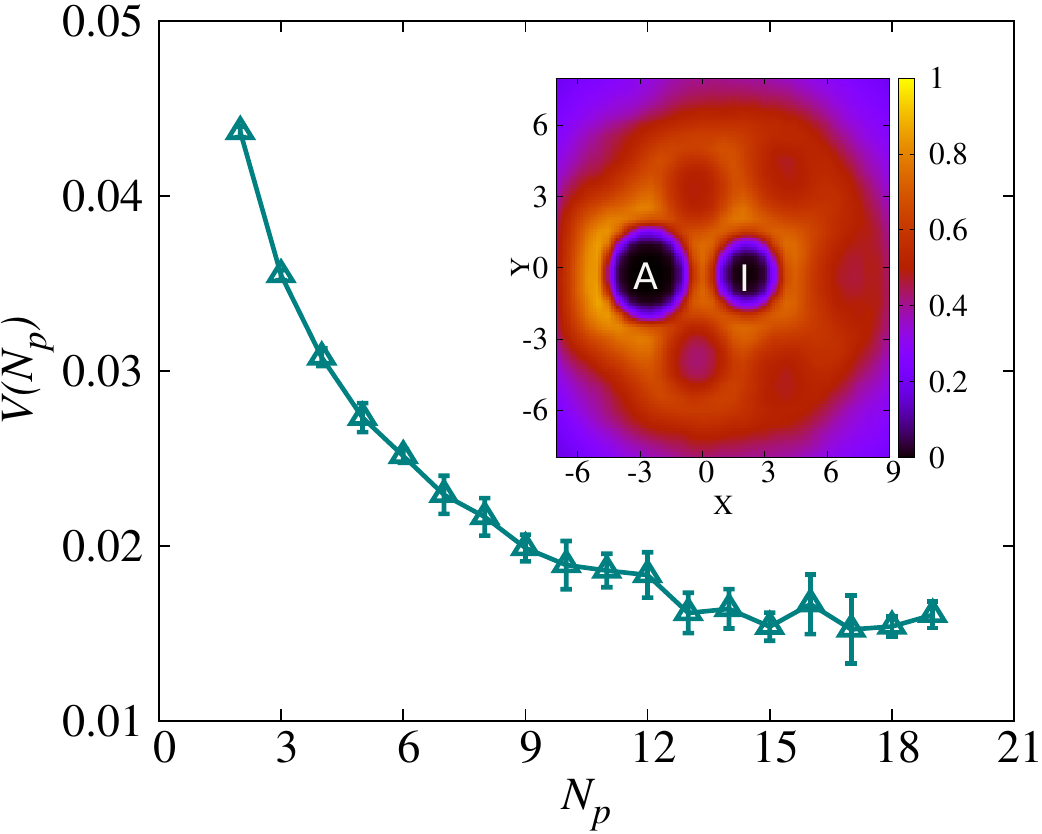}
      \caption{The variation of average speed of a cluster with cluster size $N_p$. Inset: The distribution of product solvent ($P$)  around a dimer, which is a part of a cluster with $N_p=5$.}
     \label{Fig:28_dimer_velocity}
\end{figure}

It is important to emphasize here that once the aggregates are formed  the propulsion efficiency of clusters starts to decay. To quantify this, we compute the mean-squared displacement of a single dimer with $N_d = 1$ and $28$, and average cluster speed $V(N_p)$ as a function of cluster size ($N_p$). Figure~\ref{Fig:28_dimer_msd} displays the MSD of a dimer in both kinds of fluid. Apparently, the dynamics of the dimer on average gets slower in presence of  many dimers, which is consistent for  both medium. It is noteworthy that dynamics of single as well as pair dimer is faster in case of viscoelastic fluid than the Newtonian, which is analogous to our previous observations~\cite{soudamini2019}. However, in the case of many dimers, the MSD is relatively slower for viscoelastic fluid. The slow down of  dynamics  can also be  seen in Fig.~\ref{Fig:28_dimer_velocity} in terms of the cluster speed. Here, $V(N_p) = |{\bf V}_{cmv}^{cl}|$ and ${\bf V}_{cmv}^{cl}  = \frac{ \sum_{i=1}^{N_p} {\bf V}_{i,cmv}}{N_p}$  is the center-of-mass velocity of the cluster of size $N_p$. 
Figure~\ref{Fig:28_dimer_velocity} shows that the speed of cluster decreases as the cluster size grows, and it reaches an asymptotic value beyond $N_p>12$. 
In the larger cluster, all the inert spheres aggregate together, thereby suppressing the  concentration asymmetry around the cluster resulting in the slower average speed. The concentration of the  product particles ($P$) around a dimer which is part of a cluster is shown in the inset of Fig.~\ref{Fig:28_dimer_velocity}. It is evident from the color-map that the  concentration symmetry around the dimer has been re-established by the virtue of clustering, and hence  a significant reduction in cluster speed as displayed in Fig.~\ref{Fig:28_dimer_velocity}.

To conclude this part, we observe a drastic enhancement in the aggregation of active dimers induced by the fluid elasticity. Such clustering re-enforces the concentration symmetry in the aggregate that eventually leads to slow motility of the dimers.

 \section{Conclusions}
 \label{conclusion}
 
We have presented a systematic study of the dynamics of chemically active motors with the help of a coarse-grained hybrid MD-MPC simulation in the viscoelastic and Newtonian fluids.  The motors are modelled as sphere-dimers with one end as a catalytic sphere which consumes fuel and generate product particles, whereas the other end responds to the generated product gradient thereby propelling the dimer. 
We have shown an enhancement in both translational and rotational motion of a single dimer in the presence of viscoelasticity. This is attributed to the higher concentration asymmetry and the microstructural rearrangement in the vicinity of the dimer.


Further, we have presented the assembly and dynamics of a pair-dimer.  The pair-dimer assembles  in anti-parallel  and crossed  configurations with much higher probabilities in viscoelastic fluid.  The spontaneous assembly of pair-dimer results in the diffusive motion  for anti-parallel configuration while the cross configuration still exhibits the self-propulsive dynamics with nearly same speed as the single-dimer. The observed self-assembly and its dynamics can be understood in terms of the inter-dimer long-range attraction induced by the self-generated concentration gradient. In the case of multiple dimers, we have demonstrated kinetics of cluster formation  exhibiting higher probabilities in viscoelastic fluid. Such aggregation however, results in slower translational dynamics of the cluster, which can be linked to the loss of fluid-asymmetry around the dimers, required for propulsion. 

In summary, our work emphasized the relationship between the micro-structural  relaxation of the fluid to the motility, self-assembly and the other dynamical responses of the active colloids. The present work will be helpful in providing insights of the dynamics for various types of microswimmers driven by concentration, temperature or electric field gradients in a complex medium. In general self-propelled bodies encounter several types of complex environments during their course of motion~\cite{patteson2015running,tung2017,qin2015flagella}. Understanding the impact of the complex surrounding  media on their dynamics is an exciting domain of the recent research society. A plausible extension of the present work could be the investigation of effect of viscoelasticity on microswimmers in the presence of other external factors like fluid flow, effect of confinement, and mixture of active and passive colloids.

\begin{acknowledgments}
 The computational work was performed at the HPC facility in IISER Bhopal, India. ST and SPS (grant no: YSS/2015/000230)  acknowledge SERB, DST for funding.
\end{acknowledgments}

{\bf Data Availability:}
The data that supports the findings of this study are available within the article.


\end{document}